\documentclass[a4paper,11pt]{article}
\usepackage{jheppub} 
\usepackage{lineno}
\usepackage{booktabs}
\usepackage{amsmath,amssymb,amsthm}

\newtheorem{theorem}{Theorem}[section]

\newcommand{\R}{\mathbb{R}}

\newcommand{\E}{\mathbb{E}}
\newcommand{\lmin}{\lambda_{\mathrm{min}}}

\title{What Neural Network Field Theory Can and Cannot Realise on a Computer}

\author{Thomas R. Harvey}
\affiliation{The NSF AI Institute for Artificial Intelligence and Fundamental Interactions.}
\affiliation{Center for Theoretical Physics, Massachusetts Institute of Technology, 77 Massachusetts Avenue, Cambridge, MA 02139, USA.}

\emailAdd{trharvey@mit.edu}

\abstract{One aim of neural network field theory is to put a quantum or effective field theory on a computer, with the network ensemble itself as the theory. We ask how far that aim can be pushed for a function class regular enough to be computed with. Our main result is a no-go theorem with assumptions that hold for standard network architectures. We use it to separate four versions of neural network field theory, according to whether the defining object is the finite width ensemble or its infinite width limit, and whether the target we want to compute is a quantum or an effective field theory. Neither finite width interpretation is straightforwardly consistent. For finite width ensembles with finite variance at each point, the QFT interpretation fails reflection positivity, while the EFT interpretation establishes no scale separation by which the positivity violation can be placed outside its domain of validity. Of the two limit versions, one can be simulated in full and the other only in part, as only its smeared correlators are computable with a controlled error. As such, at the level of a controlled numerical computation, the QFT and EFT versions cannot be distinguished. One dimension escapes the obstruction, yet reflection positivity is shown to still fail there at every finite width for the cosine network. Two escapes from the theorem remain, giving up either finite variance at a point or exact rotation invariance, and we discuss both of these possibilities.}

\begin{document}
\maketitle
\flushbottom

\section{Introduction}
\label{sec:intro}
Neural Network Field Theory (NN-FT)\footnote{See~\cite{Halverson:2026pmb} for an introduction.} begins with the observation that, under certain assumptions, ensembles of neural networks at infinite width behave as Gaussian processes~\cite{Halverson:2020trp,Halverson:2021aot, Halverson:2026pmb, neal1995bayesian,williams1998computing,Lee:2017qzq,matthews2018gaussian}. Concretely, let $f_\theta : \mathbb{R}^d \to \mathbb{R}$ be a neural network architecture with parameters $\theta \in \mathbb{R}^M$ drawn from a density $P(\theta)$, and $N$ (which is typically $<M$) denotes the width, or in more general architectures it should be understood as the effective number of independent contributions entering the Gaussian limit. Expectation values of the network outputs define correlation functions
\begin{equation}
    G^{(n)}(x_1,\dots,x_n) = \int d^M\theta \, P(\theta) \,
    f_\theta(x_1)\cdots f_\theta(x_n),
\end{equation}
and from the field theorist's perspective, at infinite width the ensemble over functions looks like an (in general non-local) Euclidean free field theory,
\begin{equation}
    \int d^M\theta \, P(\theta) \xrightarrow{N\to\infty} \int \mathcal{D}f \,
    e^{-S[f]},
\end{equation}
in the sense that both sides compute the same correlation functions. At infinite width the theory is free, with two-point function given by the neural network Gaussian process (NNGP) kernel
\begin{equation}
    K(x,x') = \int d^M\theta \, P(\theta) \, f_\theta(x) f_\theta(x') = \langle f(x )f(x')\rangle,
\end{equation}
so that the action is
\begin{equation}
    S[f] = \frac{1}{2}\int d^dx \, d^dx' \, f(x) \, K^{-1}(x,x') \, f(x'),
\end{equation}
where the inverse kernel is defined by
\begin{equation}
    \int d^dy \, K^{-1}(x,y) \, K(y,x') = \delta^{(d)}(x-x').
\end{equation}
The non-locality of the action stems from the fact that $K^{-1}(x,x')$ is not in general a differential operator. Only for special choices of architecture and parameter distribution does it reduce to a local form such as $-\nabla^2 + m^2$. At finite width, connected higher-point functions are suppressed in $1/N$, which are also generically non-local.

A lot of work has been done in the forward direction, where one starts with a choice of architecture, and demonstrate that properties of desired quantum field theories, including locality, conformal symmetry, and supersymmetry, can be engineered at infinite width~\cite{Demirtas:2023fir, Maiti:2021fpy, Frank:2025zuk, Frank:2026bui, Halverson:2024axc, Howard:2024bbi, Howard:2024kfd, Dogra:2026hfa}. In particular, a realisation of the free scalar at cut-off $\Lambda$ and infinite width was presented in~\cite{Halverson:2021aot, Demirtas:2023fir}.

One overarching aim of this work is to obtain a new method for simulating a quantum field theory on a computer. This paper considers difficulties with this aim, and points to where one needs to be careful about what can be claimed and computed. There is also a distinct, non-computational aim, in which the correspondence is used as a way of defining or constructing field theories, independent of whether anything is ever evaluated~\cite{Halverson:2021aot,Halverson:2026pmb}. With our aim being computation, one should also distinguish a quite different use of machine learning in this area, where a network is used as a sampler for a measure that is already defined and positive, for instance a normalising flow proposing lattice configurations for a fixed Wilson action~\cite{Albergo2019,Kanwar2020,Boyda2020,Cranmer2023,Gerdes:2022eve,Gerdes:2024rjk,Kanwar:2024ujc,SciPostPhysLectNotes.110}. While this is conceptually distinct from what has usually been called NN-FT, the underlying lattice field theory nevertheless admits an NN representation under the broader definition used in the recent universality result~\cite{Ferko:2026axm}. Here we focus on constructions closer to those of the earlier NN-FT literature~\cite{Halverson:2020trp,Halverson:2021aot,Demirtas:2023fir}.

A central difficulty with our computational aim is the inclusion of interactions. The Gaussian process results above imply that the infinite width ensemble is a free field theory. This result, however, rests on the central limit theorem applied to the final-layer neurons, so interactions may be generated by violating its assumptions~\cite{Demirtas:2023fir}. One option is to work at finite width $N$ (as is in any case required for numerical computation), where interactions arise as $1/N$ effects. An Edgeworth expansion can establish their matching order by order in $1/N$, but does not by itself establish matching to the full finite width theory, including effects non-perturbative in $1/N$. Alternatively, one can break the statistical independence of the parameters. In either case, the resulting interactions must reproduce perturbative and non-perturbative contributions together, since the neural network side of the correspondence does not distinguish between them. Any comparison with a target field theory must therefore account for both, rather than only its perturbative expansion.

For the computational task considered here, the existence of a representation that matches the desired correlators, with arbitrary architecture and parameter count, is not sufficient. The observables of interest must be computable from a finite representation, with errors that can be quantified from the ensemble and its samples without comparison with a previously known answer. We should note that actually evaluating any correlator computationally requires a Monte Carlo average over parameter draws $\theta\sim P$, whose statistical error, when well controlled, shrinks as more draws are taken. Unless otherwise stated, we assume throughout that the number of samples is not a bottleneck. This assumption becomes non-trivial when independent draws from the target measure are not directly available and must instead be generated by a sampling procedure whose convergence to the target distribution must itself be established. In that case, reliable error bars require not only control of the statistical error within the sampled distribution, but theoretical control that the sampler adequately explores the relevant sectors, including non-perturbative ones. Increasing the number of samples can reduce the former without revealing a bias from failing to sample the latter. In particular, any computational proposal should come with a way of quantifying its error from the samples and the ensemble alone, rather than relying on comparison with a known answer.

This work is distinct from the universality result of~\cite{Ferko:2026axm}, which proves that every field theory admits a neural network representation within a sufficiently broad class, with countably infinitely many parameters. Here we ask a different question: whether such a representation can be made amenable to finite computation, with controlled errors for the observables of interest\footnote{One effect of finite truncation was recently studied in~\cite{Zhang:2026tss}, which showed that architectural choices invisible at infinite width can dramatically alter finite width errors.}. This also leads to the inverse question mentioned above. Given a target QFT (or EFT), can one choose a finite computational architecture that reproduces the desired observables with errors that can be quantified without prior knowledge of their values?

It is helpful to put NN-FT into a field theorist's notation,
\begin{equation}
    Z[J] =  \int \mathcal{D}f \, \mu[f] \, e^{\int J\cdot f},
\end{equation}
where the induced field-space measure is
\begin{equation}
    \mu[f] =\int d^M\theta \, P(\theta)
    \delta[f-f_\theta].
\end{equation}
If we integrate out $f$, we are left with the push-forward of $P(\theta)$ onto the space of functions,
\begin{equation}
    \mu = F_\ast P, \qquad F:\theta \mapsto f_\theta,
\end{equation}
which does not depend on the path integral at all. A physicist reading $\int \mathcal{D}f \, e^{-S[f]}$ implicitly assumes a Gibbs density, an action, and, usually, a perturbation theory given by that density. In this case none of the three is guaranteed to exist, so the $1/N$ series is not obviously the expansion of one. An Edgeworth expansion can reconstruct an effective action order by order in $1/N$, in that their correlators, perturbative in $1/N$, agree. This perturbative action should not, however, be identified with an exact density for the finite width push-forward measure. The two agree order by order in $1/N$ without defining the same probability measure. Furthermore, a mismatch quickly comes into view. A Euclidean propagator in $d\ge2$ typically diverges at coincident points, while a typical network output has finite variance there by construction, so the two disagree about regularity before we try to engineer any other desired properties.

Before proceeding, we should separate four distinct possibilities for what NN-FT can aim to realise computationally, organised into the $2\times 2$ grid of Table~\ref{tab:grid}. The two columns differ on what the defining object is. In the finite-$N$ column the theory is the finite width iid ensemble itself. $N$ is a parameter of the theory, rather than some truncation approximation, and the $O(1/N)$ non-Gaussianities are responsible for interactions. In the limit column the theory is the $N\to\infty$ limit. In this case, finite $N$ enters only as a truncation approximation to the limiting theory, and so the $1/N$ corrections are the error of truncating the limit. Any interactions must survive the limit, which forces them to come from breaking the statistical independence of the parameters instead. The two rows differ on the target theory that we want to compute. The QFT case asks the ensemble, or its limit, to satisfy the axioms of a Euclidean quantum field theory, engineered by the choice of architecture and distribution. The EFT case asks only that there be an effective field theory, valid up to corrections that are themselves quantified, with the cut-off $\Lambda$ as a dimensionful regulator. These four possibilities are distinct and all discussed in the literature. The approach that engineers Euclidean-invariant ensembles with a tunable two-point function and reads the finite width non-Gaussianities as interactions is in the finite-$N$ column~\cite{Halverson:2021aot,Demirtas:2023fir}, while taking the infinite width limit as a definition of a field theory is the setting of the universality result~\cite{Ferko:2026axm}, which reaches a general theory only by leaving the finite-variance class. That work records mechanisms for reflection positivity in one dimension~\cite{Ferko:2025ogz} and poses it in $d\ge2$ as an open problem, and Theorem~\ref{thm:main} answers it for the finite-variance class considered here. Those one-dimensional mechanisms lie outside the finite-variance finite width class, whereas Theorem~\ref{thm:atomic} shows that reflection positivity nonetheless fails within that class in one dimension.

\begin{table}[t]
\centering
\renewcommand{\arraystretch}{1.4}
\begin{tabular}{@{}p{0.13\textwidth}p{0.40\textwidth}p{0.40\textwidth}@{}}
\toprule
& \textbf{Finite $N$}
& \textbf{The limit} \\[2pt]
& \emph{the defining object is the finite width iid ensemble; $N$ is a
parameter of the theory, and the $O(1/N)$ non-Gaussianities are its
interactions}
& \emph{the defining object is the $N\to\infty$ limit; finite $N$ is a
truncation approximation, the $1/N$ corrections are truncation error, and
interactions come from the breaking of independence} \\
\midrule
\textbf{QFT}
& Each finite width ensemble \emph{is} a Euclidean field theory: the axioms
are engineered by the choice of architecture and prior, and the $1/N$
non-Gaussianities are its interactions, exact rather than truncated.
& The limit \emph{is} the theory QFT. Finite $N$ just approximates it,
Gaussian and free for independent parameters, interacting and non-Gaussian
when the parameters are correlated. \\[4pt]
\textbf{EFT}
& Each fixed $N$ is its own \emph{exact} effective theory, with interactions
of size $1/N$; changing $N$ changes the theory, and nothing is truncated.
& The limit \emph{is} the theory EFT. The $1/N$ corrections are
a truncation error about it, and the physics is given by whatever survives the limit. \\
\bottomrule
\end{tabular}
\caption{The four possible versions of NN-FT. The columns differ on the defining object (the finite width ensemble itself, or its infinite width limit). The rows differ on the target theory (an axiomatic quantum field theory, or an effective field theory with the cut-off $\Lambda$).}
\label{tab:grid}
\end{table}

The four cases differ in how suitable they are for computation, and this paper discusses each separately. Section~\ref{sec:obs} establishes the obstruction, Section~\ref{sec:Consequences} discusses its consequences for each of the four cases in $d\ge2$, Section~\ref{sec:d=1} considers the one dimensional case, where the obstruction is absent yet reflection positivity still fails at every finite width for the cosine network, Section~\ref{sec:Ways out} returns to the assumptions of the obstruction, and finally Section~\ref{sec:Conc} concludes.

While this work was being completed, we became aware of the results of Ferko and Mutchler~\cite{Ferko:2026}, who derive complementary obstructions to finite width neural network realisations of quantum theories. While we focus on the computational interpretation of NN-FT, they study the structural limitations of finite width neural network representations of quantum theories.

\section{The Obstruction}
\label{sec:obs}

We begin by returning to the mismatch flagged in the introduction. For the propagating scalar theories of interest here, a Euclidean propagator in $d\geq2$ is typically singular at coincident points. For a free field of mass $m$,
\begin{equation}
    G(x,x) = \int \frac{d^dp}{(2\pi)^d}\,\frac{1}{p^2+m^2},
    \label{eq:pole}
\end{equation}
which is logarithmically divergent in $d=2$ and power-like in higher $d$. This coincident-point singularity is a manifestation of the distributional nature of the continuum field: the field need not exist as a random variable at a point, while suitably smeared observables remain well defined~\cite{glimm1981quantum}. Typical network ensembles are significantly simpler, having finite variance at every point,
\begin{equation}
    K(x,x) = \int d^M\theta\,P(\theta)\,f_\theta(x)^2 < \infty,
    \label{eq:finitevar}
\end{equation}
by construction\footnote{Here ``finite variance at a point'' means that $f(x)$ is an ordinary $L^2$ random variable, equivalently that its coincident connected two-point function is finite. This is stronger than what is required in a continuum QFT in $d\geq2$, where the field is generally distribution-valued and only smeared observables need have finite variance. ``By construction'' also includes the parameter distribution. A sufficiently heavy-tailed prior on the final-layer weights can make \eqref{eq:finitevar} diverge even when every individual network output is finite. In that case the divergence need not be confined to coincident points and cannot in general be removed by smearing, so conventional finite-variance Monte-Carlo error estimates need not exist. We distinguish this from the useful distribution-valued limit in Section~\ref{sec:Ways out}.}, at every width. As such, we are trying to construct an object that need not have a value at a point from one whose pointwise variance is finite.

We call a probability measure $\mu$ on functions $\R^d\to\R$ a \emph{pointwise ensemble} if $f(x)\in L^2(\mu)$ for every $x$, and write $S_2(x,y)=\E_\mu[f(x)f(y)]$. Finite-variance network ensembles form one particular subclass. Whether the $N\to\infty$ limit remains inside the pointwise class is architecture dependent. Throughout, $f$ is a single real scalar. For a multiplet of fields transforming as scalars, the theorem still applies by focusing on a diagonal component of the matrix two-point function.

A pointwise ensemble with finite variance provides a natural starting point for our computational aim. Here, individual field values are ordinary random variables for which conventional sampling errors can be assigned. In Section~\ref{sec:Ways out} we therefore consider what remains computationally accessible when this assumption is abandoned.

The property that decides whether such a Euclidean ensemble is a quantum theory is reflection positivity. This is the Osterwalder--Schrader axiom that lets one reconstruct a Hilbert space on which the Euclidean correlators become genuine quantum expectation values~\cite{osterwalder1973axioms}. Concretely, fix the hyperplane $\{x^0=0\}$ and write $\Theta$ for reflection in it, $\Theta(x^0,\vec x)=(-x^0,\vec x)$. Let $V$ be a space of observables (not necessarily restricted to monomials), evaluated in the $x^0>0$ region. Reflection positivity is then positive semi-definiteness of the matrix
\begin{equation}
    A_{uv} \;=\; \E_\mu\big[(\Theta u)\,v\big], \qquad u,v \in V .
    \label{eq:osform}
\end{equation}
This $A$ is exactly the inner product from which the physical Hilbert space is reconstructed. It will be useful to quantify this failure to be reflection positive by defining the \emph{defect}
\begin{equation}
    \delta \;=\; \max\big(0,\,-\lmin(A)\big),
    \label{eq:defect}
\end{equation}
where $\lmin(A)$ is the most negative eigenvalue of $A$. This means that $\delta=0$ is a necessary condition for a consistent quantum theory to exist on $V$.

Our obstruction only needs a necessary condition from reflection positivity. Restrict $V$ to be spanned only by the field evaluations themselves (i.e.\ degree one monomials in field evaluations), so that $A$ is the reflected two-point matrix, and its positive semi-definiteness reads
\begin{equation}
    \sum_{a,b} \bar c_a\, c_b\, S_2(\Theta x_a, x_b) \;\ge\; 0
    \qquad\text{for all } c_a\in\mathbb{C}.
    \label{eq:rp2}
\end{equation}
We call this \emph{two-point} reflection positivity, requirement (R4) below. With all this in hand, the proof of the following theorem is in Appendix~\ref{app:proof}.

\begin{theorem}
\label{thm:main}
Let $d\ge2$ and let $\mu$ be a probability measure on functions $\R^d\to\R$ such that
\begin{itemize}
  \item[\rm (R1)] $\mu$ is a pointwise ensemble: $f(x)\in L^2(\mu)$ for every $x$.
  \item[\rm (R2)] The two-point function is Euclidean invariant $S_2(gx,gy)=S_2(x,y)$ for every $g$ (translation, rotation, reflection).
  \item[\rm (R3)] $S_2$ is continuous, equivalently $f$ is mean-square continuous.
  \item[\rm (R4)] \eqref{eq:rp2} holds across a reflection hyperplane.
\end{itemize}
Then, $S_2(x,y)$ does not depend on $x$ or $y$. The field does not fluctuate from point to point, so it carries no propagating degrees of freedom and the theory it defines is trivial.\footnote{The fact that a field with constant two-point function is trivial is known~\cite{glimm1981quantum,osterwalder1973axioms}.}
\end{theorem}

Theorem~\ref{thm:main} says that in $d\ge2$, for every non-constant, Euclidean-invariant, mean-square-continuous pointwise ensemble, there is a finite configuration on which the two-point restriction of the Osterwalder--Schrader form has a negative eigenvalue. When the infinite width limit remains pointwise, this implies
\begin{equation}
    \lmin(A_\infty^{(2)}) < 0 ,
    \label{eq:floor}
\end{equation}
where $A_\infty^{(2)}$ denotes this two-point restriction. Applied instead at finite width, whenever \eqref{eq:finitevar} holds, the same argument gives $\lmin(A^{(2)}(N))<0$ at every $N$. For the usual iid construction the two-point function, and hence $A^{(2)}(N)$, can be independent of $N$. The full OS form $A(N)$ on a larger space of observables still receives finite width corrections through the higher-point functions. For independent features these may be organised as
\begin{equation}
    \lmin\big(A(N)\big) \;=\; \underbrace{\lmin(A_\infty)}_{\text{infinite width value}} \;+\; \underbrace{O(1/N)}_{\text{finite width effects}}.
    \label{eq:central}
\end{equation}
When independence is broken we cannot consider this expansion.

\section{Consequences of the Obstruction in $d\ge2$}
\label{sec:Consequences}

We now consider what the theorem implies for each of the four possibilities in the grid. We begin with ensembles engineered to satisfy (R1)--(R3) with a non-constant kernel. The theorem then forces (R4) to fail, and the four cases differ in what that means. Since this is the situation for the existing architectures in the literature, it is the natural starting point. We consider instead giving up one of the other requirements in Section~\ref{sec:Ways out}.

\subsection{(QFT\,$|$\,finite $N$): the ensemble you sample is a quantum field theory}

The claim in this possibility is that $\delta=0$ for the ensemble one actually samples, and with the $O(1/N)$ non-Gaussianities as its interactions.

A finite width network satisfying \eqref{eq:finitevar} is a pointwise ensemble at every width. If our pointwise ensemble is also designed to be Euclidean invariant with a continuous, non-constant two-point function, then Theorem~\ref{thm:main} applies to it \emph{directly} at each $N$, and reflection positivity fails. There is no reflection-positive measure behind the correlators, and so no Hilbert space and nothing that Wick-rotates to a quantum theory. 

\subsection{(QFT\,$|$\,the limit): the infinite width object is a quantum field theory}
Here the claim is $\delta=0$ at infinite width, with finite $N$ demoted to a truncation of an object that is itself the field theory. Any interaction must then survive the limit, which rules out the $1/N$ non-Gaussianities and forces it to come from correlating the parameters instead.

This case depends on whether the limit stays pointwise, and our theorem applies when it does, leading to the same conclusion as (QFT\,$|$\,finite $N$). As long as the coincident variance stays finite the limit is a pointwise ensemble, and if it is also engineered to be Euclidean invariant, continuous and non-constant, Theorem~\ref{thm:main} applies to it and forces $S_2$ constant. 

If the remaining assumptions are retained, the only way out is to leave the pointwise class by letting the coincident variance diverge in the limit. Such a limit is considered in Section~\ref{sec:Ways out}. The outcome is that it can be simulated only in part. Smeared correlators are computable with controlled error, though at finite accuracy this is indistinguishable from computing an effective theory. The singular composite operators that would distinguish the limit as a genuine quantum field theory cannot be computed with errors controlled from the ensemble alone. For the purposes of a controlled computation, this option is therefore indistinguishable from (EFT\,$|$\,the limit).

\subsection{(EFT\,$|$\,finite $N$): the ensemble you sample is an effective field theory}

The claim here is that each fixed $N$ defines its own effective theory, with the $O(1/N)$ terms interpreted as genuine interactions, while a cut-off $\Lambda$ sets the scale below which the description is valid. Changing $N$ therefore changes the theory rather than merely improving an approximation.

Theorem~\ref{thm:main} shows that reflection positivity fails at every finite $N$, but this does not by itself rule out an EFT interpretation. An EFT only needs to reproduce physical observables below its cut-off, and a violation that requires momenta of order $\Lambda$ is allowed. The intended interactions and the reflection-positivity defect both arise at finite width, but it is not clear that they have identical dependence on the independent ratio $p/\Lambda$. The ensemble could define a physical EFT if, for observables with $p\ll\Lambda$,
\[
    \text{interaction}
    \sim \frac{1}{N},
    \qquad
    \text{RP violation}
    \sim
    \frac{1}{N^\alpha}
    \left(\frac{p}{\Lambda}\right)^k,
    \qquad k>0, \qquad \alpha\geq 0.
\]
The intended interaction would then remain visible at order $1/N$, while the positivity violation would lie parametrically below it within the EFT's domain of validity. Theorem~\ref{thm:main} does not exclude this possibility, since it establishes the existence of a negative direction without determining the scale at which that direction appears.

Despite not being ruled out, such a separation may be difficult to achieve in practice. The required separation must be engineered and demonstrated on the intended low-energy observables. Moreover, a perturbative calculation in $1/N$ need not capture every finite width effect. For example, reflection-positivity violations inaccessible analytically, including contributions non-perturbative in $1/N$ such as $e^{-cN}$, could still be present in the full numerical ensemble. If the separation is absent, the reflection-positivity violation remains visible at some order of the coupling. Conversely, if all finite width effects are placed below the claimed accuracy, the $O(1/N)$ interactions have also become noise rather than resolved physics. We therefore do not obtain a strict no-go theorem for the (EFT$\,|\,$finite $N$) interpretation, but believe it unlikely that all of these effects can be brought systematically under control.

\subsection{(EFT\,$|$\,the limit): the infinite width object is an effective field theory}

Here the $1/N$ corrections are, again, the error of truncating the limit, and the only regulator left carrying physics is the cut-off $\Lambda$.

A regulated propagator at cut-off $\Lambda$ has a finite coincident variance and is Euclidean invariant, so Theorem~\ref{thm:main} applies to it. $\delta\neq0$ at every finite $\Lambda$, already at infinite width~\cite{Bailleul2025}. This is not by itself a problem for an effective theory, which only claims agreement with a target theory below $\Lambda$ and need not require the regulated object to be reflection positive at all scales. The coincident variance whose finiteness was fatal in the QFT case probes distances below $\sim1/\Lambda$, outside the regime of validity of the EFT. Therefore, when using NN-FT to simulate (EFT$\,|\,$the limit), one requires, observable by observable, that the finite-$N$ truncation error be pushed below the systematic error already imposed by $\Lambda$. This requirement includes finite width effects non-perturbative in $1/N$, which are invisible to the perturbative expansion discussed in the previous section.

\section{One Dimension}
\label{sec:d=1}

In one dimension the obstruction is absent, as the momentum integral \eqref{eq:pole} converges, and the free propagator is finite at coincidence, $G(x,x)=1/(2m)$, so the infinite width limit can be a legitimate reflection-positive field without diverging. In $d\ge2$ the theorem ruled out (QFT\,$|$\,finite $N$). Although the obstruction no longer applies, reflection positivity can still fail at finite width, as we now demonstrate for the cosine network.

Throughout this section the ensemble is a sum of $N$ iid features, $f=c_N\sum_{i=1}^N g_i$ with the $g_i$ independent and identically distributed. This applies to any architecture where the final layer features can be considered independent. We assume also that the coincident variance is held finite and width-independent, which for iid features fixes the normalisation $c_N\propto N^{-1/2}$. This is the structure of many constructions in the literature, including the realisation of the free field~\cite{Halverson:2021aot,Demirtas:2023fir}.

Take the target theory to carry finitely many masses, so that its two-point function $K$, the kernel any candidate ensemble must realise, is a positive combination of finitely many decaying exponentials, equivalently a bounded, non-constant, completely monotone function.\footnote{The free scalar of mass $m$ has two-point function $\tfrac{1}{2m}e^{-m|t|}$. Up to the normalisation fixed by $K(0)$ this is the single-mass case $K(t)=e^{-m|t|}$.} For such a $K$, $A_\infty$ is positive semi-definite and the infinite width Gaussian theory is reflection positive~\cite{glimm1981quantum}. For finitely many masses,
\[
K(t)=\sum_{q=1}^{r}\alpha_q e^{-m_q|t|},
\]
and so the reflected two-point matrix is
\[
(A_\infty)_{ij}
=
K(t_i+t_j)
=
\sum_{q=1}^{r}
\alpha_q e^{-m_q t_i}e^{-m_q t_j}.
\]
Once more than $r$ distinct times are included, $A_\infty$ must have a null direction, so
\begin{equation}
    \lmin(A_\infty)=0.
    \label{eq:zerofloor}
\end{equation}
This contrasts with the strictly negative value \eqref{eq:floor} in $d\ge2$ for a non-trivial pointwise limit. In one dimension, the infinite width theory is therefore reflection positive but degenerate, and any reflection-positivity defect must come entirely from finite width corrections.

With our normalisation, the two-point function is independent of $N$, so a defect can appear only with correlators involving more insertions. As shown in Appendix~\ref{app:d1}, on the span of monomials of degree at most three, the Osterwalder--Schrader form terminates
\begin{equation}
    A(N) \;=\; A_\infty + \frac{D}{N} + \frac{E}{N^2},
    \label{eq:termination}
\end{equation}
with $A_\infty$ fixed by the target and $D$, $E$ model-dependent, set by the higher single-feature moments.

As these possible defects are model dependent, as a concrete example we take the cosine network,\footnote{The random-cosine, or random-Fourier-feature, has a long history~\cite{Campbell1909,Rice1944,Wiener1923}.}
\begin{equation}
    f(x) \;=\; \sqrt{\frac{2K(0)}{N}}\,\sum_{i=1}^{N}\cos\big(w_i x + b_i\big),
    \qquad w_i \stackrel{\text{iid}}{\sim}\rho,\quad b_i\stackrel{\text{iid}}{\sim}\mathrm{U}[0,2\pi],
    \label{eq:cosnet}
\end{equation}
where $\rho$ is a distribution of your choice, and the model is exactly stationary at every finite width. That is, its two-point function depends on the two inputs only through their difference, which in one dimension is equivalent to Euclidean invariance. By Bochner's theorem\footnote{Bochner's theorem: a continuous function $K$ is positive definite if and only if it is the Fourier transform of a finite positive measure, $K=\hat\rho$. Normalising $K(0)=1$ makes $\rho$ a probability distribution, which is what allows it to serve as the distribution of the network's weights.\label{fn:bochner}} the distribution $\rho$ realises every stationary kernel, the completely monotone kernels of the free theories among them. At infinite width the Gaussian limit is determined entirely by its kernel. Since the cosine network realises every stationary kernel through the choice of $\rho$, specialising to this architecture still includes every one-dimensional free theory, local or non-local. We now truncate this infinite width limit, and interpret the $1/N$ corrections as interactions. In other words, the theorem below addresses only the (QFT\,$|$\,finite $N$) version of NN-FT.

The following theorem is proven in Appendix~\ref{app:d1}.
\begin{theorem}
\label{thm:atomic}
Let the target theory be bounded and completely monotone, with finitely many masses,
\[
K(t)=\sum_{q=1}^{r}\alpha_q\,e^{-m_q|t|}
\]
where $\alpha_q>0$, $\sum_q\alpha_q=K(0)$, and at least one $m_q>0$. Then reflection positivity of the cosine network \eqref{eq:cosnet} fails at \emph{every} finite width $N$. The defect is at least of order $1/N^2$, and of order $1/N$ whenever $\ker A_\infty$ contains a direction $v$ with $v^{\mathsf T}Dv<0$.
\end{theorem}

Combined with Section~\ref{sec:Consequences}, (QFT\,$|$\,finite $N$) is therefore false in every dimension, in $d\ge2$ for every typical architecture and in $d=1$ for at least the cosine network with a finite number of masses.

For example, $\rho(w)=\frac{m}{\pi(w^2+m^2)}$ gives $K(t)=e^{-m|t|}$, a natural finite width realisation of the free-field target of~\cite{Halverson:2021aot,Demirtas:2023fir}. Although each individual cosine feature is smooth, the heavy-tailed frequency distribution reproduces the non-differentiable free-field two-point function exactly at every finite width, so the obstruction is not simply a consequence of the smoothness of the network. Its two-point function has the desired free-field form exactly, at every width, and yet the full finite width measure violates reflection positivity for every finite $N$. This defect is computed in Appendix~\ref{app:d1}.

\section{Two Ways Forward}
\label{sec:Ways out}
So far we have taken Theorem~\ref{thm:main} as given and, in Section~\ref{sec:Consequences}, considered its consequences when its requirements are met, and commented on the consequences of keeping reflection positivity exact. We now ask whether relaxing any of the other requirements leaves a situation that is amenable to computation. The two escapes worth considering are relaxing either finite variance at a point (R1) or exact rotation invariance (R2).

\subsection{Escape one: give up finite variance at a point.}
Letting the coincident variance diverge takes the ensemble out of the pointwise class. This does not necessarily prevent the computation of some observables, but one must identify the observable sector for which finite-variance estimators and controlled errors survive. While we will not discuss how such a limit might be engineered, we do consider the computational difficulties that arise if it can be achieved. The coincident variance is also the variance of a single Monte-Carlo draw of the field at a point, so the same divergence that evades the theorem prevents the limiting field from being sampled there as a finite-variance random variable. This allows a genuine interacting theory to appear in an infinite width limit without contradicting the theorem, the Liouville construction~\cite{David2016} being a known example, and is consistent with the universality result of~\cite{Ferko:2026axm}, which allows the target field theory to be distribution-valued.

The divergence is confined to coincident points and therefore does not necessarily affect smeared observables. A smeared observable $O[\varphi]=\int d^dx\,\varphi(x)\,O(x)$, with $\varphi$ a fixed test function, has estimator variance $\iint\varphi\,\varphi\,\langle OO\rangle$, which is finite exactly when the two-point function of $O$ has an integrable singularity. We call such observables safe. For a safe observable, error bars shrink as more Monte-Carlo draws are added, and increasing $N$ to approach the limit is a legitimate convergence check that requires no prior knowledge of the full answer. This condition assumes that the divergence is confined to coincident points, so that $\langle OO\rangle$ remains well defined at separated points. When the divergence instead enters at the feature level, as for heavy-tailed final-layer weights, it remains at separated points and cannot in general be removed by smearing. The variance is then divergent and no finite error bar can be assigned. 

For observables that do admit error bars, computing them to finite accuracy is indistinguishable from working in the (EFT$\,|\,$the limit) approach of Section~\ref{sec:Consequences}, since finite resolution implicitly introduces a smoothing scale that hides the coincident-point divergence.

For QFT, we also need to consider the singular composite operators, built from the field at a coincident point, so their self-correlator $\langle OO\rangle$ diverges at coincidence. The Liouville vertex operators $:e^{\alpha f}:$ are an example. Normal ordering removes the singularity, and so the expectation is $O(1)$, but a single-draw estimate has variance $\mathcal O\!\left(e^{\alpha^2 K(0)}\right)$ as the cut-off $\Lambda$ is removed and $K(0)\to\infty$. In $d=2$ the coincident variance grows logarithmically, $K(0)\sim\log\Lambda$, so the estimator variance is a power $\Lambda^{\alpha^2}$ of the cut-off. This becomes exponentially expensive, $\sim e^{\Lambda^{d-2}}$, for $d\ge3$.\footnote{The heavy tail is a property of the operator, not the network. A singular composite is built from the field at a coincident point, so in an ordinary lattice simulation the same operator has an equally heavy-tailed estimator, its coincident $\langle OO\rangle$ diverging as the spacing is removed. This is a signal to noise issue, that is also present in lattice simulations, and its exponentially hard cases lie beyond naive sampling there as well. Although the lattice does not reduce this cost, the locality of its measure can be used to improve this situation.}

A reliable computation, one whose errors can be established without prior knowledge of the answer, needs a UV resolution scale external to the ensemble, and the difficulty is removing it, since singular composites demand ever more ultraviolet information as the resolution approaches the continuum. Finite width is not such a scale, since $N\to\infty$ does not refine a cut-off at fixed theory in the way that the lattice spacing does. The lattice also has local sampling, and with it variance-reduction tools that the parameter-space measure does not naively provide~\cite{Luscher:2001up}. Therefore finite width sampling controls the smeared sector at finite resolution, but provides no trusted continuum errors for singular composites. The Liouville three-point function of~\cite{Ferko:2026axm} illustrates the issue. Its match to the DOZZ formula holds only up to an overall constant, since truncation renormalises the coupling, and its importance sampling is validated by agreement with the known answer, no longer covering the exact result near the Seiberg bound. This is a valid match to a known interacting theory, but establishing its accuracy required information external to the computation itself. This becomes a problem if the same methodology is to be used to compute properties of field theories for which no analytic answer is known, which is ultimately one of the primary purposes of numerical simulation.

\subsection{Escape two: give up exact rotation invariance.}
This option keeps finite variance and drops (R2). The lattice lives in this category, as does one dimension since there are no rotations to average over in Step~3 of the proof. Under the broad definition of neural network representation used in the universality result of~\cite{Ferko:2026axm}, the lattice itself is included, as its finite-dimensional field configuration can be regarded as the parameter space, with the map to the field configuration as the architecture. Therefore the lattice provides an example in which this escape is realised with a controlled dimensionful parameter. However, as noted earlier, we focus here on architectures closer to those considered in the earlier NN-FT literature.

Conventional NN-FT architectures can also drop (R2). For example, consider $d$ independent copies of the cosine network \eqref{eq:cosnet}, one per coordinate, and multiply them. The two-point function is the product kernel
\[
    K(x-y)=\prod_{i=1}^{d} e^{-m|x_i-y_i|},
\]
which has finite variance at every point, is continuous and non-constant, and is invariant only under a discrete subgroup of the full rotation group. It nonetheless satisfies two-point reflection positivity (R4) across every coordinate hyperplane. At the level of two-point functions, then, such an architecture can reproduce the lattice's discrete rotation group.

The important difference is that this construction has no analogue of the lattice spacing $a$ controlling the symmetry breaking. On the lattice, $a$ is an independent length that can be sent to zero, with the residual anisotropy removed order by order through Symanzik's programme~\cite{Symanzik1983a,Symanzik1983b}.\footnote{Improved (Symanzik) lattice actions trade exact finite-$a$ reflection positivity for improved rotational invariance, while the Wilson action is the reflection-positive case and accepts $O(a)$ symmetry breaking. The escape proposed here targets the Wilson-style case, since Theorem~\ref{thm:main} excludes exact Euclidean invariance together with the other assumptions in the pointwise class.} By contrast, the dimensionful hyperparameters of the NN-FT constructions considered here, such as the mass scale appearing in the weight distribution, determine the target kernel rather than independently controlling its anisotropy. They therefore provide no parameter with which the rotation breaking can be systematically removed while holding the target theory fixed. This is a limitation of these constructions, not of neural network representations in the broad sense of~\cite{Ferko:2026axm}. Whether the more conventional NN-FT architectures considered here can realise such a controlled continuum limit remains an open question.

\section{Conclusion}
\label{sec:Conc}

Most of this paper explores the consequences of Theorem~\ref{thm:main}. In $d\ge2$ any pointwise, Euclidean-invariant, mean-square-continuous ensemble that is reflection positive across one hyperplane has a constant two-point function, and so is trivial. This is independent of any higher correlators.

The strongest statement the paper reaches concerns (QFT\,$|$\,finite $N$). For the finite-variance class considered by Theorem~\ref{thm:main}, a finite width ensemble cannot be an exact reflection-positive quantum field theory in $d\ge2$. In one dimension the theorem does not apply, and yet Theorem~\ref{thm:atomic} shows that reflection positivity still fails at every finite width for the cosine network, with a finite number of masses. The obstruction is therefore not finite width itself, or a lack of expressive power. In $d\ge2$ it is the simultaneous demand for finite variance at each point, exact Euclidean invariance, mean-square continuity, and reflection positivity for a non-constant two-point function.

(QFT\,$|$\,the limit) can be computed only in part. Its smeared correlators are reachable with controlled error, but computing them to finite accuracy is indistinguishable from computing a regulated effective theory. The singular observables that distinguish the continuum theory require separate control of their estimator errors as the regulator is removed, which is not supplied by the finite width approximation itself.

(EFT\,$|$\,finite $N$) is not established as a consistent interacting theory. Reflection positivity fails at every width within the class considered, but this need not invalidate an EFT if the violation can be parametrically separated from the intended low-energy interactions. No such separation follows from the $1/N$ expansion itself, and it would have to be engineered through an independent scale dependence and demonstrated on the intended low-energy observables, while also controlling finite width effects not visible perturbatively. Such a regime appears very difficult to engineer.

What remains is (EFT\,$|$\,the limit). A network is a legitimate way to compute the correlators of an effective field theory defined and regulated at infinite width, with the finite width ensemble a truncation of the limit and all finite width errors, including those non-perturbative in $1/N$, pushed below the systematic errors already imposed by the cut-off.

This paper also explored two possible escapes from the primary theorem, by relaxing its assumptions. Giving up finite variance at a point reaches genuine interacting limits, Liouville theory among them, but shifts the computational question to which observables retain finite-variance estimators and controlled errors as the regulator is removed. Giving up exact rotation invariance keeps finite variance, with the lattice providing a controlled example that is itself included under the broad definition of neural network representation used in~\cite{Ferko:2026axm}. For architectures closer to those of the earlier NN-FT literature, the constructions considered here can reproduce the lattice's discrete symmetry at the two-point level, but lack its independent dimensionful parameter controlling the anisotropy. Building such an architecture with a controlled continuum limit, and asking whether the theorem relaxes for fields of higher spin, are questions we leave open.

\acknowledgments
The author would like to thank Christian Ferko, James Halverson and Mathis Gerdes for reading an early version of the draft, Jessica Howard for insightful discussions, and Siddharth Mishra-Sharma for providing a Claude Max subscription. The author is supported by the National Science Foundation under Cooperative Agreement PHY-2019786 (The NSF Institute for Artificial Intelligence and Fundamental Interactions, \url{http://iaifi.org/}). The author acknowledges the use of AI for proofreading and theorem development. All AI-assisted content was independently checked by the author, who takes full responsibility for the accuracy of the work.

\appendix

\section{Proof of Theorem~\ref{thm:main}}
\label{app:proof}

Throughout, $\mu$ is a probability measure on the space of functions $\R^d\to\R$, specified by its finite-dimensional distributions, and (R1)--(R4) are as in Section~\ref{sec:obs}:
\begin{itemize}
  \item[(R1)] $f(x)\in L^2(\mu)$ for every $x$, with $S_2(x,y)=\E_\mu[f(x)f(y)]$ and $S_2(x,x)<\infty$;
  \item[(R2)] the two-point function is Euclidean invariant: $S_2(gx,gy)=S_2(x,y)$ for every rigid motion $g$ (translation, rotation, reflection);
  \item[(R3)] $S_2$ is continuous, equivalently $f$ is mean-square continuous;
  \item[(R4)] the two-point reflection positivity \eqref{eq:rp2} holds across one hyperplane.
\end{itemize}
The proof can be summarised by: Reflection positivity makes the correlator convex along the reflection axis. Rotation invariance in $d \geq 2$ makes its slope at the origin vanish, so convexity forces it to be non-decreasing. A bounded convex function with vanishing slope at the origin is constant.

\paragraph{Step 1: momentum-space representation.}

By (R2), the two-point function is translation invariant, so
\[
S_2(x,y)=K(x-y),
\]
where $K$ is real, rotation invariant, and positive definite, which allows us to rewrite (R1) in terms of $K(0)$ as
\[
K(0)=S_2(x,x)<\infty,
\]
so the pointwise variance is finite, while Cauchy--Schwarz implies
\[
|K(x)|\le K(0)
\]
for all $x$. By (R3), $K$ is continuous.

A continuous positive-definite function admits a momentum-space representation. By Bochner's theorem (footnote~\ref{fn:bochner}), there exists a \emph{finite} positive measure $\nu$ on momentum space such that
\begin{equation}
    K(x)=\int e^{ip\cdot x}\,d\nu(p), \qquad \nu(\R^d)=K(0).
\end{equation}
Therefore finite pointwise variance is equivalent to requiring that the total contribution from all momentum modes is finite. Since $K$ is rotation invariant, uniqueness in Bochner's theorem implies that $\nu$ is rotation invariant as well.

\paragraph{Step 2: reflection positivity implies convexity.}

In this step, we demonstrate that reflection positivity implies the resulting two-point function, viewed as a function of Euclidean time, is necessarily convex.

To isolate the dependence on Euclidean time, we smear the field in the $d-1$ directions parallel to the reflection plane using a fixed time-independent and spatially (i.e. not including Euclidean time) rotationally symmetric test function $\varphi$. In momentum space this simply multiplies the measure by the positive weight
\[
w(\mathbf p)=|\hat\varphi(\mathbf p)|^2\ge0,
\]
which depends only on $|\mathbf p|$ because $\varphi$ is rotationally symmetric. The smeared two-point function at Euclidean time separation $t$ is
\begin{equation}
    k_w(t)
    :=
    \iint
    \varphi(\mathbf x)\varphi(\mathbf y)\,
    K(t,\mathbf x-\mathbf y)\,
    d\mathbf x\,d\mathbf y
    =
    \int
    e^{ip_0t}\,
    w(\mathbf p)\,
    d\nu(p)
    =
    \int_{\mathbb R}
    \cos(p_0t)\,
    d\lambda_w(p_0),
    \label{eq:kwdef}
\end{equation}
where the second equality follows from the momentum-space representation of $\varphi$ and $K$ of Step~1. Integrating over the transverse momenta produces a finite positive measure $\lambda_w$ for the $p_0$. Since $\nu$ is reflection invariant, $\lambda_w$ is even, allowing the exponential to be replaced by the cosine.

Applying (R4) to the difference of two smeared time slices at $t_1,t_2>0$ gives
\[
\iint
\beta(s)\beta(t)\,
k_w(s+t)\,
ds\,dt
\ge0,
\]
where $\beta=\beta_1-\beta_2$ and $\beta_1,\beta_2$ are supported near $t_1,t_2$, respectively.

Finally, concentrating $\beta_1$ and $\beta_2$ to delta functions at $t_1,t_2$, again using the continuity of $k_w$ inherited from (R3), yields
\begin{equation}
    k_w(2t_1)
    -2k_w(t_1+t_2)
    +k_w(2t_2)
    \ge0,
    \qquad
    t_1,t_2>0.
    \label{eq:convex}
\end{equation}
Therefore $k_w$ is midpoint convex. Since $k_w$ is continuous, midpoint convexity is equivalent to convexity, so every spatially smeared Euclidean two-point function is convex as a function of Euclidean time.

\paragraph{Step 3: rotational invariance makes the initial slope finite.}

We now show where the assumption $d\ge2$ enters. By \eqref{eq:kwdef}, the right derivative of $k_w$ at the origin is finite provided that the smeared Euclidean-energy distribution has a finite first moment,
\begin{equation}
    \int |p_0|\,w(\mathbf p)\,d\nu(p)<\infty.
    \label{eq:firstmoment}
\end{equation}
Since $\nu$ is invariant under the full rotation group, its contribution at fixed momentum magnitude $r=|p|$ may be replaced by its angular average over the sphere. Spatial smearing suppresses modes with large transverse momentum $|\mathbf p|$. Consequently, on a sphere of large radius $r$, only narrow polar caps near $p_0=\pm r$ contribute appreciably.

To make this precise, write
\[
    p_0=r\cos\vartheta,
    \qquad
    |\mathbf p|=r\sin\vartheta,
\]
and let
\[
    c_d=\int_0^\pi \sin^{d-2}\vartheta\,d\vartheta.
\]
The angular average of $|p_0|w(\mathbf p)$ at fixed radius $r$ is
\begin{equation}
    I(r)
    =
    \frac{2}{c_d}\,
    r\int_0^1 w(rs)\,s^{d-2}\,ds,
    \label{eq:angular-average}
\end{equation}
where $s=\sin\vartheta$.

As $\varphi$ is a Schwartz test function\footnote{A Schwartz test function is a smooth function whose derivatives decay faster than any power at infinity. Its Fourier transform is again Schwartz, so for every $M\ge1$ there exists $C_M>0$ such that
\[
|\hat\varphi(\mathbf p)|\le C_M(1+|\mathbf p|)^{-M}.
\]}, so is $w(\mathbf p)=|\hat\varphi(\mathbf p)|^2$. At large $r$, substituting
$u=rs$ in \eqref{eq:angular-average} gives
\begin{equation}
    I(r)
    \le
    A_d\,r^{2-d},
    \qquad
    A_d
    =
    \frac{2}{c_d}
    \int_0^\infty w(u)\,u^{d-2}\,du.
    \label{eq:I-large-r}
\end{equation}
At small $r$, the bound $w(\mathbf p)\le\|w\|_\infty =\sup_{\mathbf p}w(\mathbf p)$ gives
\begin{equation}
    I(r)
    \le
    B_d\,r,
    \qquad
    B_d
    =
    \frac{2\|w\|_\infty}{c_d(d-1)}.
    \label{eq:I-small-r}
\end{equation}
The second bound controls $I(r)$ for small $r$, while the first controls it for large $r$ (remaining constant for $d=2$ and decaying for $d>2$). Together they give a uniform bound on $I(r)$
\[
    \sup_{r\ge0} I(r)<\infty.
\]
Using rotation invariance to replace the integrand by its angular average,
we therefore obtain
\begin{equation}
    \int |p_0|\,w(\mathbf p)\,d\nu(p)
    \le
    \bigl(\sup_r I(r)\bigr)\nu(\mathbb R^d)
    <\infty,
\end{equation}
where the final inequality uses the finiteness of the momentum-space measure established in Step~1.

We can get some geometric intuition for why we have this dimension requirement. On a sphere of radius $r$, spatial smearing leaves only polar caps whose relative area scales as $r^{-(d-1)}$. Inside those caps, $|p_0|\sim r$, so their total
contribution scales as
\[
    r\,r^{-(d-1)}=r^{2-d}.
\]
This remains bounded precisely for $d\ge2$. In one dimension there are no transverse directions to smear.

\paragraph{Step 4: convexity and a finite initial slope force the spectrum to zero momentum.}

Write $c(t)=k_w(t)$, and take the time derivative
\[
c'(t)
=
-\int p_0\sin(p_0t)\,d\lambda_w(p_0),
\]
which is continuous and satisfies
\[
c'(0)=0.
\]
By \eqref{eq:convex}, $c$ is convex, so $c'$ is non-decreasing on $(0,\infty)$, implying
\[
c'(t)\ge c'(0)=0,
\]
and therefore $c$ is non-decreasing on $[0,\infty)$.

On the other hand,
\[
c(t)\le c(0),
\]
since $c$ is a two-point function and is bounded above by its value at the origin. The only non-decreasing function that never exceeds its initial value is a constant. Thus
\[
c(t)=c(0).
\]
Using \eqref{eq:kwdef}, this gives
\[
\int \bigl(1-\cos(p_0t)\bigr)\,d\lambda_w(p_0)=0
\qquad
\text{for all }t.
\]
The integrand is non-negative, and for every $p_0\neq0$ it is positive for some $t$, so $\lambda_w$ is supported entirely at $p_0=0$.

Convexity alone is not enough here. The crucial input is the finite initial slope, established in Step~3, which holds only for $d\ge2$.

\paragraph{Step 5: assembly.}

Let $\varphi$ be a Gaussian, so $w>0$ everywhere. By Step~4, the projected measure $\lambda_w$ is supported at $p_0=0$. Since $w(\mathbf p)>0$ for all $\mathbf p$, this is possible only if $\nu$ itself is supported on the hyperplane $\{p_0=0\}$.

The choice of reflection direction was arbitrary. Repeating the argument after rotating the reflection plane shows that $\nu$ is supported on every rotated hyperplane $R\{p_0=0\}$. For $d\ge2$, the intersection of all such hyperplanes is the single point $\{0\}$. Hence $\nu$ is supported entirely at zero momentum, so $K$ is constant. Therefore
\[
\E\big[(f(x)-f(y))^2\big]
=
2\bigl(K(0)-K(x-y)\bigr)
=
0
\]
for all $x,y$. Thus $f(x)=f(y)$ almost surely, so the field is spatially constant and the theory is trivial. \hfill$\square$

\section{Reflection positivity fails at every finite width ($d=1$) for the cosine network}
\label{app:d1}

This appendix proves Theorem~\ref{thm:atomic}, and related material. Let $f$ be the cosine network \eqref{eq:cosnet}, let $V$ denote the space of monomials of degree at most three in its half-space evaluations, and let $A(N)$ be the corresponding Osterwalder--Schrader form.

\paragraph{The two-point function is exact at every width.} For the cosine network \eqref{eq:cosnet} the two-point function is
\begin{equation}
\begin{split}
    S_2(x,y) &=\frac{2K(0)}{N}\sum_{i,j}\E\big[\cos(w_ix+b_i)\cos(w_jy+b_j)\big] \\
    &=\frac{2K(0)}{N}\sum_{i}\tfrac12\,\E_w\big[\cos(w(x-y))\big]
     =K(0)\,\E_w\big[\cos(w(x-y))\big]=K(0)\,\hat\rho(x-y),
\end{split}
    \label{eq:cos2pt}
\end{equation}
where the cross terms $i\neq j$ vanish after averaging over $b$. It carries no $N$, so the degree-one block of $A(N)$ receives no finite width correction.

\paragraph{The $1/N$ expansion is exact on $V$.}
Write the ensemble as $f=N^{-1/2}\sum_i g_i$ with independent mean-zero features $g_i$, so that $g_i(x)=\sqrt{2K(0)}\cos(w_ix+b_i)$ for the cosine network. An entry of $A(N)$ on $V$ is a moment containing at most six field insertions. Since the bias is uniformly distributed, every odd single-feature moment vanishes. Consequently all entries between monomials of opposite degree parity vanish, so $A(N)$ is block diagonal between the even and odd sectors. The reflection-positivity defect therefore lives entirely in the odd sector, and it is enough to consider the degree-one and degree-three monomials.

The remaining entries are even moments $\E[f(s_1)\cdots f(s_{2r})]$ with $r\le3$. Expanding $f=N^{-1/2}\sum_i g_i$ produces a sum over neuron labels. Independence, and mean-zero, means that only terms in which equal labels are grouped together survive. Equivalently, the surviving contributions are indexed by partitions of the $2r$ insertions into blocks, each block representing insertions coming from the same neuron. A block of size $b$ contributes the single-feature moment $M_b=\E\prod_{a=1}^b g(s_a)$, while singleton blocks vanish because $\E g=0$. This gives
\begin{equation}
    \E\prod_{k=1}^{2r} f(s_k)
    \;=\; N^{-r}\sum_{\text{partitions }\pi}\,(N)_{|\pi|}\prod_{B\in\pi}M_{|B|}(s_B),
    \qquad (N)_k=N(N-1)\cdots(N-k+1),
    \label{eq:moments}
\end{equation}
the sum running over partitions $\pi$ of the $2r$ slots into blocks of sizes $|B|$, with $(N)_{|\pi|}$ counting the ways to place the $|\pi|$ blocks on distinct neurons. The prefactor $N^{-r}(N)_{|\pi|}$ is a polynomial in $1/N$ of degree $r-1$, so on $V$ the series stops at $1/N^2$. This is the exact termination \eqref{eq:termination}. The expansion \eqref{eq:moments} used neither one dimension nor a particular activation, only that the features are iid with finite moments up to the sixth, so it holds at points of $\R^d$ and for any single-layer network, and with it the termination at $1/N^2$ on $V$. The parity argument asks for one thing more, the vanishing of the odd single-feature moments, which the uniform bias supplies for any activation odd about its mean and not only the cosine.

\paragraph{The leading defect is controlled by $D$.} With $\lmin(A_\infty)=0$ from \eqref{eq:zerofloor} and the exact termination \eqref{eq:termination}, it is enough to examine $\ker A_\infty$. For $v\in\ker A_\infty$,
\begin{equation}
    v^{\mathsf T}A(N)v
    =
    \frac{v^{\mathsf T}Dv}{N}
    +
    \frac{v^{\mathsf T}Ev}{N^2}.
    \label{eq:kernelsign}
\end{equation}
Therefore the presence or absence of an $O(1/N)$ defect is determined entirely by the model-dependent matrix $D$. If $v^{\mathsf T}Dv<0$ for some $v\in\ker A_\infty$, reflection positivity already fails at order $1/N$.
Otherwise the leading contribution is $O(1/N^2)$. The proof below follows the latter route, showing first that the diagonal entry vanishes and then identifying an off-diagonal coupling through $D$ that forces negativity.

\subsection{Proof of Theorem~\ref{thm:atomic}}
We now prove Theorem~\ref{thm:atomic}. Write $K(t)=\sum_{q=1}^{r}\alpha_q\,e^{-m_q|t|}$ with $\alpha_q>0$, $\sum_q\alpha_q=K(0)$, and at least one $m_q>0$. We also normalise $K(0)$ to $1$ without loss of generality. The proof proceeds in four steps. Step~1 finds a null direction of the infinite width Osterwalder--Schrader form. Step~2 shows that the exact two-point function makes its diagonal $1/N$ correction vanish. Step~3 shows that finite width couples it to a cubic observable. Step~4 shows that this coupling makes the resulting two-dimensional restriction indefinite, establishing reflection-positivity violation at every finite width.
\paragraph{Step 1: a null direction at infinite width.}

Take $r+1$ distinct positive times
\[
0<t_1<\cdots<t_{r+1}.
\]
The degree-one$\times$degree-one block of $A_\infty$ is the reflected two-point matrix
\[
H_{pp'}
=
K(t_p+t_{p'})
=
\sum_{q=1}^{r}
\alpha_q\,e^{-m_qt_p}e^{-m_qt_{p'}},
\]
a sum of $r$ rank-one matrices,\footnote{Its entries depend on the two times only through their sum, so $H$ is a Hankel matrix, but only the rank bound is used.} and therefore of rank at most $r$. Since there are $r+1$ distinct times, $H$ has a one-dimensional kernel because the vectors $u_q$ are linearly independent.

Let $c$ span this kernel. Then
\[
0
=
c^{\mathsf T}Hc
=
\sum_q
\alpha_q
\left(
\sum_pc_pe^{-m_qt_p}
\right)^2.
\]
Since every $\alpha_q$ is strictly positive, each term in the sum must vanish separately:
\[
\sum_pc_pe^{-m_qt_p}=0
\qquad
\text{for every }q.
\]
We refer to these identities as the \emph{kernel condition}.

Now set
\[
v=\sum_pc_pf(t_p).
\]
Then $v\in\ker A_\infty$. The degree-one$\times$degree-one block is annihilated by construction. The degree-one$\times$degree-three block also vanishes. At infinite width the ensemble is Gaussian, so Wick's theorem expresses every contraction in
\[
\mathbb E[(\Theta f(t_p))f(t_a)f(t_b)f(t_c)]
\]
as a sum of terms containing a factor $K(t_p+t_x)$, and the kernel condition makes each such factor vanish,
mass by mass.

\paragraph{Step 2: the width correction cannot affect the two-point data.}
The two-point function is exact at every width by \eqref{eq:cos2pt}, so the degree-one block of $A(N)$ carries no $1/N$ correction and $v^{\mathsf T}Dv=0$, for our $v$ and indeed for every degree-one direction.

\paragraph{Step 3: finite width couples the null direction to a degree-three observable.}

Take the degree-three observable
\[
\psi=f(t_1)^2f(T),
\qquad
T>t_{r+1}+2t_1.
\]
The quantity $v^{\mathsf T}D\psi$ is the $1/N$ contribution to $\E[(\Theta v)\psi]$. By \eqref{eq:moments} it is the connected single-feature four-point function,
\[
\sum_p c_p
\left[
M_4(-t_p,t_1,t_1,T)
-
\sum_{\text{pairings}}M_2M_2
\right].
\]

Write each cosine as $\cos u=\tfrac12(e^{iu}+e^{-iu})$. The uniform bias average removes every term with non-zero overall bias phase. The surviving terms are
\begin{equation}
    M_4(-t_p,t_1,t_1,T)
    =
    \frac{K(0)}{2}
    \Big[
        2K(t_p+T)
        +
        K(T-t_p-2t_1)
    \Big],
    \label{eq:appM4}
\end{equation}
while the disconnected pairings contribute
\[
2\,K(t_p+t_1)\,K(T-t_1)
+
K(0)\,K(t_p+T).
\]

Every term containing $K(t_p+t_x)$ is set to zero by the kernel condition. The only surviving contribution is $K(T-t_p-2t_1)$. Since $T>t_{r+1}+2t_1$, its argument is positive, so
\[
K(T-t_p-2t_1)
=
\sum_q
\alpha_q
e^{-m_q(T-2t_1)}
e^{+m_qt_p},
\]
which is not annihilated by the kernel condition, due to the wrong sign in the exponent. With $K(0)=\sum_q\alpha_q=1$, this leaves
\begin{equation}
    v^{\mathsf T}D\psi
    =
    \frac12
    \sum_q
    \alpha_q
    e^{-m_q(T-2t_1)}
    \hat g(m_q),
    \qquad
    \hat g(m)
    =
    \sum_p
    c_p
    e^{+mt_p}.
    \label{eq:appcoupling}
\end{equation}

If some $m_q=0$, its contribution vanishes because $\hat g(0)=\sum_pc_p=0$, which is the kernel condition at zero mass. Writing $h(z)=\sum_pc_pe^{-zt_p}$, so that $\hat g(m)=h(-m)$, the kernel condition says that $h(m_q)=0$ for every mass. These are all the real zeros of $h$,\footnote{A non-trivial linear combination of $k$ exponentials with distinct exponents has at most $k-1$ real zeros~\cite{KarlinStudden1966}. Here $h$ is a combination of the $r+1$ exponentials $e^{-zt_p}$, so it has at most $r$ real zeros. The kernel condition already supplies the $r$ zeros $m_q$, and so no others remain.} and therefore $\hat g(m_q)=h(-m_q)\neq0$ for every positive mass. Therefore \eqref{eq:appcoupling} is a non-trivial finite exponential sum in $T$, and such a sum has only finitely many zeros. Choosing $T$ away from them gives
\[
v^{\mathsf T}D\psi\neq0.
\]

\paragraph{Step 4: the two-dimensional restriction is indefinite.}

Restrict the Osterwalder--Schrader form to $\mathrm{span}\{v,\psi\}$. Since $A_\infty v=0$, $v^{\mathsf T}Dv=0$, and $E$ is supported entirely on the degree-three block, so that $v^{\mathsf T}Ev=v^{\mathsf T}E\psi=0$, the restricted matrix is
\[
\begin{pmatrix}
0 & v^{\mathsf T}D\psi/N\\[2mm]
v^{\mathsf T}D\psi/N & \psi^{\mathsf T}A(N)\psi
\end{pmatrix}.
\]
Its determinant is
\[
-\left(\frac{v^{\mathsf T}D\psi}{N}\right)^2<0
\]
at every finite width. A symmetric $2\times2$ matrix with negative determinant has one positive and one negative eigenvalue. Reflection positivity therefore fails for every finite width.

The entry $\psi^{\mathsf T}A(N)\psi$ affects only the magnitude of the negative eigenvalue, never its existence. If instead this entry vanished, the defect would scale as $1/N$ instead. This establishes the unconditional statement of Theorem~\ref{thm:atomic}.

\paragraph{The $1/N$ term, and the free field.}

If, in addition, $\ker A_\infty$ contains a direction with $v^{\mathsf T}Dv<0$, then \eqref{eq:kernelsign} gives
\[
v^{\mathsf T}A(N)v<0
\]
for every
\[
N>\max\!\left(0,\frac{v^{\mathsf T}Ev}{|v^{\mathsf T}Dv|}\right),
\]
so the defect becomes order $1/N$. Any such direction necessarily mixes degree-one and degree-three monomials, since $v^{\mathsf T}Dv=0$ on the degree-one block alone.

The proof relies on a null direction of the infinite width Osterwalder--Schrader form. Such a direction exists when the target has finitely many masses, but for a representing measure of infinite support the reflected two-point matrix may be strictly positive definite on every finite collection of times. The construction therefore does not extend directly, and whether the theorem remains true in that setting is open.
\hfill$\square$


\bibliographystyle{JHEP}
\bibliography{biblio}






\end{document}